\documentclass[twocolumn,showpacs,preprintnumbers,amsmath,amssymb]{revtex4}
\usepackage[english]{babel}
\usepackage{dcolumn}
\usepackage{bm}

\begin{document}

\title{Comments on paper by B.Lesche (Phys. Rev. E {\bf 70}, 017102 (2004))\\
"Renyi entropies and observables" }

\author{A. G. Bashkirov}
\email{abas@idg.chph.ras.ru}

\affiliation{Institute Dynamics of Geospheres, Russian Academy of
Sciences, Moscow, Russia} 

\date{\today}
 \maketitle

In 1982 Leshe published paper \cite{Lesche} on instability of the
Renyi entropy. His paper was developed later by Abe \cite{Abe}.
From the time of their publications they are  often referred as a
mortal verdict for the Renyi entropy.

This problem became the subject of our discussion with Abe
\cite{BashSt,Abe2}

Another attack on the Renyi entropy was recently launched by
Lesche \cite{Lesche2}. The most important points of his paper
related to the subject are outlined in the following paragraphs.

He considers firstly an initial macrostate that characterizes by
eigenvalues $ a_I ,b_I , . . . ,h_I$ of macroscopic commeasurable
observables $A,B,...,H$. The special characteristics of this kind
of state are the following: 1) The number of occupied microstates
$W_I=W_{a_I ,b_I , . . . ,h_I}$ is of the order $M^N$ and
therefore any individual probability $(W_I)^{-1}$ is extremely
small (of the order of $M^{-N}$). 2) The number of empty
microstates is larger than the number of occupied microstates by a
huge factor, which is also of the order of $\tilde M^N$.
Essentially these two characteristics make this sort of state a
problem case.

This initial state is of the form
\begin{equation}
\rho=\sum_j^{W_I}|a_I ,b_I , . . . ,h_I,j\rangle\frac
1{W_I}\langle a_I ,b_I , . . . ,h_I,j|
\end{equation}
where $|a_I ,b_I , . . . ,h_I,j\rangle$ is a basis of common
eigenstates (Dirac's ket-vectors) and $j$ is an index of
degeneracy of the initial macrostate.

All Renyi entropies of this state have the same value, which is
the Boltzmann entropy of the macrostate
\begin{equation}
S_q^{(R)}=S^{(B)}=\ln W_I
\end{equation}
Now, imagine that a friend of ours enters the laboratory and
criticizes our experiment. He claims our preparation of state may
in some cases result in the macrostate $[\bar{a},\bar{b} , . . .
,\bar{h}]$ with the number of occupied microstates
$\overline{W}=W_{\bar a ,\bar b , . . . ,\bar h}$. Now, if the
probability of such cases, say, $\delta =10^{-100}$, we will not
be able to convince our friend that our probability assignment is
better than his by showing experimental results. His density
operator would be
\begin{eqnarray}
\tilde\rho&=&\sum_j^{W_I}|a_I ,b_I , . . . ,h_I,j\rangle\frac
{1-\delta}{W_I}\langle a_I ,b_I , . . .
,h_I,j|\nonumber\\&+&\sum_j^{\overline{W}}|\bar a ,\bar b , . . .
,\bar h,j\rangle\frac {\delta}{\overline{W}}\langle \bar a ,\bar b
, . . . ,\bar h,j|
\end{eqnarray}

The Renyi entropy of the friend's probability assignment is
\begin{eqnarray}
\tilde S^{(R)}_q(\delta)&=&\frac 1{1-q}\ln\Big \{(1-\delta)^q
(W_I)^{1-q} +
\delta^q (\overline{W})^{1-q}\Big\}\nonumber\\
&=&\frac 1{1-q}\ln \Big \{\delta^q
(\overline{W})^{1-q}\nonumber\\&\times &\Big (1+\frac
{(1-\delta)^q}{\delta^q}\frac {(W_I)^{1-q}}{(\overline{W})^{1-q}}
\Big )\Big\}\nonumber\\&=&\ln \left \{\delta^{\frac
{q}{1-q}}\overline{W}\right\} \nonumber\\&+ &\frac
{1}{1-q}\ln\left(1+\frac {(1-\delta)^q}{\delta^q}\frac
{(W_I)^{1-q}}{(\overline{W})^{1-q}}\right )
\end{eqnarray}
The first term is of order N. To estimate the second term we now
distinguish the following two cases: (a) If $q>1$, we shall assume
that our friend thought of a state $[\bar a ,\bar b , . . . ,\bar
h]$ with smaller entropy than the main state $[a_I ,b_I , . . .
,h_I]$, that is,
\begin{equation}
\overline{W}\ll \delta^{\frac {q}{q-1}} W_I\,\,\,{\rm for}\,\,q>1
\end{equation}
If $S^{(B)}(a_I ,b_I , . . . ,h_I)- S^{(B)}(\bar a ,\bar b , . . .
,\bar h)$ is macroscopic (of the order $N$), the third term is
clearly also negligible as compared to the first one. (b) If
$q<1$, we assume that the friend thought of a state $[\bar a ,\bar
b , . . . ,\bar h]$ whose entropy is macroscopically larger than
$S^{(B)}(a_I ,b_I , . . . ,h_I)$, that is
\begin{equation}
\overline{W}\gg \delta^{-\frac {q}{1-q}} W_I\,\,\,{\rm for}\,\,q<1
\end{equation}
Again the second term will be negligible.
\begin{equation}
\tilde S^{(R)}_q(\delta)=\ln \left \{\delta^{\frac
{q}{1-q}}\overline{W}\right\}
\end{equation}
or
\begin{equation}
\tilde S^{(R)}_q(\delta)=\ln \overline{W} + {\frac
{q}{1-q}}\ln\delta=\ln \left \{\delta^{-1}\overline{W}\right\} +
{\frac {1}{1-q}}\ln\delta
\end{equation}
Then, according to Lesche, the second terms of this equations are
negligible as compared to the first ones (for instance, with
$N\approx 10^{24}$ and $\delta=10^{-100}$ his argument is
restriction to $q$ values with $|1-q|\gg 10^{-22}$). So, in either
case, the Renyi entropy of the friend's probability assignment
would essentially be the entropy of the irrelevant state $[\bar a
,\bar b , . . . ,\bar h]$
\begin{equation}
\tilde S^{(R)}_q\simeq\ln \overline{W}\simeq \ln \left
\{\delta^{-1}\overline{W}\right\},
\end{equation}
which is far away from our initial value, Eq. (2).\\

I begin my comments with the remark that Lesche did not touche the
same problem for the Tsallis entropy in spite of that he thanks C.
Tsallis and S. Abe for bringing his attention to the subject that
he abandoned many years ago.

The Tsallis entropy of the initial state $\rho$ is
\begin{equation}
S^{(Ts)}_q=\frac 1{1-q}\left (W_I^{1-q}-1\right)
\end{equation}
For the friend's probability assignment $\tilde\rho$ it becomes
\begin{eqnarray}
\tilde S^{(Ts)}_q(\delta)&=&\frac 1{1-q} \{(1-\delta)^q
(W_I)^{1-q} +
\delta^q (\overline{W})^{1-q}-1\}\nonumber\\
&=&\frac 1{1-q} \delta^{q}(\overline{W})^{1-q}\left\{1+\frac
{(W_I)^{1-q}-1}{\delta^{q}(\overline{W})^{1-q} }\right\}
\end{eqnarray}
With the use of the same speculations that have led Lesche to Eq.
(9) the Tsallis entropy for both cases $q>1$ and $q<1$ takes the
form
\begin{equation}
\tilde S^{(Ts)}_q(\delta)=\frac 1{1-q}\left ((\delta^{\frac
{q}{1-q}}\overline{W})^{1-q}-1\right)
\end{equation}
So, we see that the Tsallis entropy is far away from its initial
value, Eq. (10), as well as the Renyi entropy.

Under restrictions (5) and (6) they both correspond to an equally
probable distribution $p_j=(\delta^{\frac
{q}{1-q}}\overline{W})^{-1}$ (for all $j$) related to the friend's
state $[\bar a ,\bar b , . . . ,\bar h]$ without regard for the
iitial state $[a_I ,b_I , . . . ,h_I]$. Just this result should be
expected. Indeed, states with greatest (least) probabilities
contribute significantly to both $q$-entropies at $q>1$ ($q<1$).
The restrictions (5) and (6) may be rewritten for the
probabilities as
\begin{equation}
\frac {\delta}{\overline{W}}\gg \frac {1-\delta}{W_I}\cdot
10^{\frac {100}{q-1}}\,\,\,{\rm for}\,\,q>1;
\end{equation}
\begin{equation}
\frac {\delta}{\overline{W}}\ll \frac {1-\delta}{W_I}\cdot
10^{-\frac {100}{1-q}}\,\,\,{\rm for}\,\,q<1
\end{equation}
Then, the resulted equations for $\tilde S^{(R)}_q(\delta)$ and
$\tilde S^{(Ts)}_q(\delta)$ are to be accepted as quite natural,
but not as evidences of their instabilities.

It is interesting to discuss an alternative form of the Renyi and
Tsallis entropies expansions.

The more natural form of the Renyi entropy expansion should start
with the Boltzmann entropy, Eq. (2), as a leading term of the
expansion. Then
\begin{eqnarray}
S^{(R)}_q(\delta)&=&\frac 1{1-q}\ln \left \{(1-\delta)^q
(W_I)^{1-q}\right.\nonumber\\&\times &\left(1+\frac
{\delta^q}{(1-\delta)^q}\frac
{(\overline{W})^{1-q}}{(W_I)^{1-q}}\left. \right
)\right\}\nonumber\\&=&\ln W_I+\frac {q}{1-q}\ln (1-\delta)
\nonumber\\&+ &\frac {1}{1-q}\ln\left(1+\frac
{\delta^q}{(1-\delta)^q}\frac
{(\overline{W})^{1-q}}{(W_I)^{1-q}}\right )
\end{eqnarray}
The second term of this expansion is negligible when $|1-q|\gg
\delta$. The third term  becomes negligible when the macrostate
proposed by the friend fulfills the inequalities
\begin{equation}
\frac {(\overline{W})^{1-q}}{(W_I)^{1-q}}\ll \delta^{-q}
\end{equation}
or, taking into consideration that $\delta=10^{-100}$,
\begin{equation}
\frac {\delta}{\overline{W}}\ll \frac {1-\delta}{W_I}\cdot
10^{\frac {100}{q-1}}\,\,\,{\rm for}\,\,q>1;
\end{equation}
\begin{equation}
\frac {\delta}{\overline{W}}\gg \frac {1-\delta}{W_I}\cdot
10^{-\frac {100}{1-q}}\,\,\,{\rm for}\,\,q<1
\end{equation}
We see that $\delta$--pre-factors in both last inequalities
alleviate restrictions imposed on $\overline{W}$, in contrast to
the role of such $\delta$--pre-factors in the inequalities (13),
(14) of the Lesche's expansion. It means that the third term of
our expansion can be neglected in much more numerous cases than
the one of the Lesche expansion. In particular, cases of
$\overline{W}\simeq W_I$ are included here.

It may be supposed that the macrostate $[\bar{a},\bar{b} , . . .
,\bar{h}]$ chosen by the Lesche's friend will fulfill inequalities
(17) or (18), but not (13) or (14) with overwhelming probability,
if only the friend is not a grandson of Maxwell's demon. So, the
Renyi entropy of the friend's probability assignment would be
essentially the same Boltzmann entropy of our state
\begin{equation}
S^{(R)}_q(\delta)\simeq\ln W_I.
\end{equation}

These arguments repel the Lesche's attack on stability of the
Renyi entropy.

The similar alternative expansion of the Tsallis entropy under the
same restrictions gives rise to
\begin{eqnarray}
\tilde S^{(Ts)}_q(\delta)&=&\frac 1{1-q} \{(1-\delta)^q
(W_I)^{1-q} +
\delta^q (\overline{W})^{1-q}-1\}\nonumber\\
&=&\frac 1{1-q}(1-\delta)^q W_I^{1-q}\Big\{1+\frac
{\overline{W}^{1-q}-1}{(1-\delta)W_I^{1-q}}\Big\}\nonumber\\
&\simeq&\frac 1{1-q}\left\{(1-\delta)^q W_I^{1-q}-1\right\}\nonumber\\
&\simeq&\frac 1{1-q}\left\{ W_I^{1-q}-1\right\}
\end{eqnarray}
Thus, we have get the Tsallis entropy of the initial state.

The above speculations count in favor of stability of both Renyi
and Tsallis entropies relative stochastic perturbations of an
initial macrostate with the uniform distribution of probabilities
of microstates.

\end{document}